\PassOptionsToPackage{nonamebreak}{natbib} 
\documentclass[a4paper,fleqn,usenatbib]{mnras}

\usepackage{newtxtext,newtxmath}
\usepackage[T1]{fontenc}
\usepackage{ae,aecompl}

\usepackage{float}
\usepackage{array}
\usepackage{relsize}
\usepackage{graphicx}	% Including figure files
\usepackage{amsmath}	% Advanced maths commands
\usepackage{amssymb}	% Extra maths symbols
\usepackage{lscape}

\def\aap{A\&A}%

\def\aj{AJ}%
\def\apj{ApJ}%
\def\apjl{ApJ}%
\def\apjs{ApJS}%
\def\ao{Appl.~Opt.}%
\def\mnras{MNRAS}%
\def\nat{Nature}%
\title[Dynamical mass estimates of NGC1052-DF2]{Effects of mass models on dynamical mass estimate: the case of ultra diffuse galaxy NGC1052-DF2}

\author[K. Hayashi \& S. Inoue]{
Kohei Hayashi$^{1}$\thanks{E-mail: kohei.hayashi@ipmu.jp} and
Shigeki Inoue$^{2,3}$
\\
$^{1}$Institute for Cosmic Ray Research~(ICRR), The University of Tokyo\\
$^{2}$Kavli Institute for Physics and Mathematics of the Universe~(Kavli IPMU), The University of Tokyo\\
$^{3}$Department of Physics, School of Science, The University of Tokyo, Bunkyo, Tokyo 113-0033, Japan
}

\date{Accepted XXX. Received YYY; in original form ZZZ}

\pubyear{2018}

\begin{document}
\label{firstpage}
\pagerange{\pageref{firstpage}--\pageref{lastpage}}
\maketitle

\begin{abstract}
NGC1052-DF2 was recently discovered as the dark-matter deficient galaxy claimed by~\citet[][vD18]{vD2018a}. 
However, large uncertainties on its dynamical mass estimate have been pointed out, concerning the paucity of sample, statistical methods and distance measurements.
In this work, we discuss the effects of the difference in modeling of the tracer profile of this galaxy on the dynamical mass estimate.
To do this, we assume that the tracer densities are modeled with power-law and S\'ersic profiles, and then we solve the spherical Jeans equation to estimate the dynamical mass. 
Applying these models to kinematic data of globular clusters in NGC1052-DF2, we compare 90 per cent upper limits of dynamical mass-to-light ratios estimated between from this analysis and from vD18.
We find that the upper limit obtained by the power-law is virtually the same as the result from vD18, whilst this limit estimated by the S\'ersic is significantly greater than that from vD18, thereby suggesting that NGC1052-DF2 can still be a dark-matter dominated system.
Consequently, we propose that dynamical mass estimate of a galaxy is largely affected by not only small kinematic sample but the choice of tracer distributions, and thus the estimated mass still remains quite uncertain.
\end{abstract}

% Select between one and six entries from the list of approved keywords.
% Don't make up new ones.

%%%%%%%%%%%%%%%%%%%%%%%%%%%%%%%%%%%%%%%%%%%%%%%%%%

%%%%%%%%%%%%%%%%% BODY OF PAPER %%%%%%%%%%%%%%%%%%
%%%%%%%%%%%%%%%%%%%%%%%%%%%%%%%%%%%%%%%%%%%%%%%%%%%%%%%%%%%%%%%%%%%%%%%%%%%%%%%5
%%%%%%%%%%%%%%%%% INTRODUCTION S1 %%%%%%%%%%%%%%%%%%
%%%%%%%%%%%%%%%%%%%%%%%%%%%%%%%%%%%%%%%%%%%%%%%%%%%%%%%%%%%%%%%%%%%%%%%%%%%%%%%5
\section{Introduction}

Owing to recent deep photometric observations, ultra diffuse galaxies~(UDGs)
have been discovered in clusters and groups of galaxies~\citep[e.g.,][]{vanetal2015a,vanetal2015b,Kodetal2015,Yagetal2016,vandetal2017,Truetal2017}.
These galaxies have commonly the typical luminosity of a dwarf galaxy, 
but they are similar to Milky-Way-sized galaxies in physical size.
Therefore, UDGs are characterized as an extremely low surface brightness galaxies.
From dynamical analysis for kinematic data of globular clusters~(GCs) within UDGs, 
they are, in general, thought to be largely dominated by dark matter as well 
as the the Galactic dwarf spheroidal galaxies~\citep[e.g.,][]{vanetal2016}, 
but how these diffuse galaxies are formed and evolved in their dark matter halo 
is still ongoing debate~\citep[e.g.,][]{vanetal2015a,vanetal2015b,AL2016,DiCetal2017}.

Interestingly enough, however, \citet[][hereafter vD18]{vD2018a} have recently 
discovered a dark matter deficient the UDG that is deficient in dark matter, 
NGC1052-DF2, which is a satellite of NGC1052~elliptical galaxy.
They adopted mass tracer estimator~(MTE) constructed by~\citet{Watetal2010} 
to estimate the dynamical mass within a given radius, and applied it to the 
kinematic data of the 10 GCs of the galaxy.
Then, they estimated the dynamical mass to be only $<3.4\times10^8M_{\odot}$
(at 90\% confidence) within 7.6~kpc from its centre, even though the stellar 
mass of this galaxy is estimated to be $2\times10^8M_{\odot}$.

If their mass estimation is correct, this UDG is a certainly exciting galaxy in 
terms of the deficit of dark matter and understanding its formation~\citep[e.g.,][]{Ogi2018}.
However, previous studies have pointed out uncertainties of this mass estimation 
due to the paucity of kinematic sample, statistical methods and distance measurements~\citep{Lapetal2018,Maretal2018,Truetal2018}.
All of them argued that the mass estimate of NGC1052-DF2 still remains largely 
uncertain, hence it is difficult to conclude that the UFD is a galaxy laking dark matter. 
  
In this paper, we point out uncertainties of tracer models assumed in dynamical mass 
estimations. In particular, we show that the dynamical mass of NGC1052-DF2 might be 
affected by tracer distribution models assumed in analysis.
As mentioned above, vD18 utilized MTE to determine the dynamical mass.
This mass estimator is based on the projected virial theorem and a spherical Jeans equation.
Moreover, this estimator assumes that the density profile of the tracers is modeled with 
single power-law form because of requirement from their analytic treatment in the MTE modelling.
However, a power-law profile is only acceptable to have a diverged profile at the centre 
of system without any apparent physical motivation or evidence.
Furthermore, since vD18 reported that the stellar system of NGC1052-DF2 is fitted with 
a S\'ersic profile, which has cored profile in inner parts, it may be natural to 
expect that GC tracers might follow a similar profile.
Therefore, in order to investigate the effects of model differences on mass 
estimate, especially tracer distribution, we calculate dynamical masses of 
NGC1052-DF2 with two models: S\'eric and power-law tracer density profiles.
In addition, to estimate dynamical mass, we do not utilize MTE but use 
line-of-sight velocity dispersion derived from the spherical Jeans equation.
Thus, we set constraints on dark halo parameters from the information of 
positions and line-of-sight velocities of tracers, and then we estimate 
the dynamical mass using these best-fitting parameters.
However, in principle, both methods should result in virtually similar results 
if there is not significant statistical uncertainty in the tracer distribution.

This Letter is organized as follows. In Section 2 we introduce the method of 
dynamical mass estimation based on our analysis.
In Section 3, we show the results of mass estimation and then comparison 
with vD18's estimation. Summary and conclusion are shown in Section 4.

%%%%%%%%%%%%%%%%%%%%%%%%%%%%%%%%%%%%%%%%%%%%%%%%%%%%%%%%%%%%%%%%%%%%%%%%%%%%%%%5
%%%%%%%%%%%%%%%%% Dynamical mass estimations S2 %%%%%%%%%%%%%%%%%%
%%%%%%%%%%%%%%%%%%%%%%%%%%%%%%%%%%%%%%%%%%%%%%%%%%%%%%%%%%%%%%%%%%%%%%%%%%%%%%%5
\section{Dynamical mass estimations}
In this section, we briefly introduce the methods of dynamical mass estimates 
for NGC1052-DF2 based on spherical Jeans equations.
Since NGC1052-DF2 is far from the Sun~($D_{\odot}\sim20$~Mpc estimated from vD18), 
the available observed information of its GCs are their projected distributions, 
line-of-sight velocities and velocity dispersions.
Thus, Jeans equations should be integrated along the line of sights.
In assumptions of spherically symmetric mass distribution and no net-streaming 
motions for the tracers, the line-of-sight velocity dispersion is straightforwardly written as
\begin{eqnarray}
\sigma^{2}_{\text{l.o.s}}(R) = \frac{2}{\Sigma_{\ast}(R)}\int^{\infty}_{R}dr
\Bigl(1-\beta_{\text{ani}}\frac{R^2}{r^2}\Bigr)\frac{\nu_{\ast}(r)\sigma^{2}_{r}(r)}{\sqrt{1-R^2/r^2}},
\label{eq:displos}
\end{eqnarray}
where $R$ denotes the projected radius from the centre of the galaxy, 
and $\Sigma_{\ast}(R)$ is the projected tracer distribution integrated
by $\nu_{\ast}(r)$ along the line-of-sight direction.
The three dimensional velocity dispersions of the tracers in the system 
are represented with $\sigma_r$,~$\sigma_{\theta}$, and $\sigma_{\phi}$ 
which denote components along radial, polar and azimuthal directions, respectively.
For spherical symmetry, we take $\sigma_{\phi}=\sigma_{\theta}$.
The anisotropy parameter $\beta_{\text{ani}}=0$ indicates an isotropic 
velocity ellipsoid of the tracers, while positive and negative $\beta_{\text{ani}}$
 are radially- and tangentially-biased velocity dispersions, respectively.  
Then, the anisotropy parameter, $\beta_{\text{ani}}$, is defined as 
$\beta_{\text{ani}}=1-\sigma_{\theta}/\sigma_r$.
Radial dispersion $\sigma_r$ is obtained by the spherical Jeans equation under 
assumptions of steady-state and dark matter dominated system~\citep{BT2008},
which is expressed as

\begin{eqnarray}
\sigma^{2}_{r}(r) = 
\frac{1}{\nu_{*} (r) }
\int^{\infty}_{r} 
\nu_{*} (r')
\left(\frac{r'}{r}\right)^{2 \beta_{\text{ani}}}
\frac{G M(r')}{r'^2} 
dr'\ ,
\label{eq:dispr}
\end{eqnarray}
where $G$ is the gravitational constant, and $M(r)$ is the enclosed mass 
of the spherical dark matter halo{:} 
$M(r) \equiv \int^{r}_{0} 4 \pi r'^2 \rho_{\text{DM}} (r') dr'$.
From Eq.\,(\ref{eq:displos}) and Eq.\,(\ref{eq:dispr}), 
we can estimate the DM profile $\rho_{\text{DM}}$ by adopting the 
line-of-sight velocity dispersion $\sigma_{\text{l.o.s}} (R)$ to the 
observational kinematic data.

% figure 1
\begin{figure}
	% To include a figure from a file named example.*
	% Allowable file formats are eps or ps if compiling using latex
	% or pdf, png, jpg if compiling using pdflatex
	\includegraphics[width=\columnwidth]{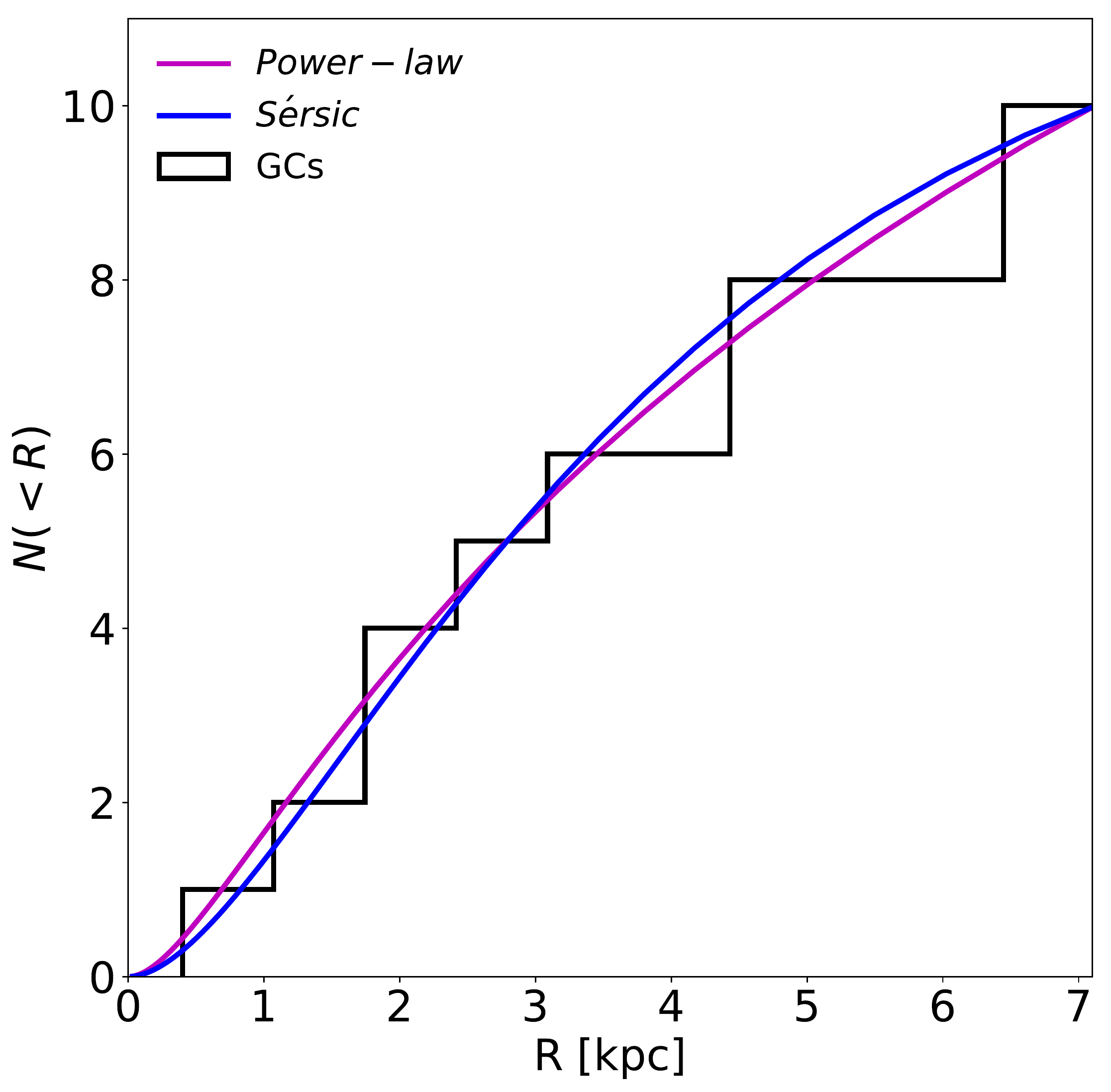}
    \caption{Projected cumulative number distribution of GC tracers in the NGC1052-DF2~(black).
    Magenta and blue curves illustrate the best-fit profiles modeled by power-law and S\'ersic profiles, respectively.}
    \label{CND}
\end{figure}

%%%%%%%%%%%%%%%%%  S2.1 %%%%%%%%%%%%%%%%%%
\subsection{Tracer number density models}
In order to solve the above Jeans equation, a three-dimensional profile 
of tracer number densities is assumed.
vD18 assumed that their number density falls off according to a power 
law, $\nu_{\ast}(r)\propto r^{-\gamma_{\ast}}$, following~\citet{Watetal2010},
and they found $\gamma_{\ast}=0.9\pm0.3$\footnote{They checked the effects
 of errors in the number density of GCs on dynamical estimates and found 
 that for more shallower slope~($\gamma_{\ast}=0.5$) the mass decreases 
 by 20~per~cent and for more steeper one~($\gamma_{\ast}=1.5$) the mass 
 increases by 30~per~cent.}.
However, there is no physical justification for whether this cusped profile 
is the most likely model, and thus there is a possibility that a cored profile is also acceptable.
 
In this work, we therefore assume that the member tracers in the NGC1052-DF2 are 
distributed according to a S\'ersic profile~\citep{Ser1968} as well as a power-law profile.
This is motivated by the facts that the stellar distribution of this galaxy is 
expressed by a two-dimensional S\'ersic profile~(vD18), and the GC tracers might 
follow a distribution similar to the stars. 
S\'ersic profile on the sky plane is written by $\Sigma_{\ast}(R)\propto\exp[-R^{1/m}]$, 
where $m$ is the S\'ersic index, which measures the curvature of the profile, 
and $m=1$ corresponds to the exponential profile, and $R$ is the projected radius from 
the centre of the galaxy. The three-dimensional density $\nu_{\ast}(r)$ is obtained 
from the surface density $\Sigma_{\ast}(R)$ by deprojection through the Abel transform derived by~\citet{Limetal1999}.

To obtain structural parameters of power-law and S\'ersic profiles, 
we fit them to the cumulative profile of the projected number density of the tracers (GCs).
Since, as was done in vD18, we employ the power-law model that is fitted to the 
three-dimensional density profile of the tracers, we use Abel integral to calculate 
the projected power-low profile, and then we cumulate the projected number density with 
$R$ from inside to outside. The distribution and the best-fitting profiles 
are shown in Figure~\ref{CND}.
We use data of the 10 GCs including the positions and line-of-sight velocities published 
by~\citet{vD2018b}, and we employ a simple $\chi^2$ fitting.
From the fitting results, we find $\gamma_{\ast}=3.16$ for the power-law model with 
$\chi^2_{\rm PL}=1.89$ and $m=1.3$ for the S\'ersic model with $\chi^2_{\rm Ser}=2.16$, respectively.
In comparison with these models, there is no significant difference 
in the goodness of fit, due to the paucity of the sample tracers.
In what follows, we calculate dynamical mass of the galaxy 
using these two tracer number density models.
When we solve the Jeans equation, we adopt two kinds of scale radii of the tracer distribution.
One is the half-light radius of the galaxy. 
vD18 determined this radius to be 2.2~kpc using deep photometric data obtained from {\it Hubble Space Telescope}.
The other one is the radius of half number of the GC tracers. This radius is also estimated by vD18 to be 3.1~kpc.
As described below, we show that the dependence of our mass estimation on the scale radii appears to be negligible.

%%%%%%%%%%%%%%%%%  S2.2 %%%%%%%%%%%%%%%%%%
\subsection{Dark matter halo model}
For the dark matter halo, we adopt a generalized Hernquist profile given by~\citet{Her1990} and \citet{Zhao1996},
\begin{eqnarray}
&& \rho_{dm}(r) = \rho_0 \Bigl(\frac{r}{r_{s}} \Bigr)^{-\gamma}\Bigl[1+\Bigl(\frac{r}{r_{s}} \Bigr)^{\alpha}\Bigr]^{-\frac{\beta-\gamma}{\alpha}},
\label{DMH}
\end{eqnarray}
where $\rho_0$ and $r_s$ are the scale density and radius, $\alpha$ is
the sharpness parameter of the transition from the inner slope $-\gamma$ to the outer slope $-\beta$. 
For $(\alpha,\beta,\gamma)=(1,3,1)$, we recover the NFW profile~~\citep[][]{NFW1996,NFW1997} motivated by cosmological pure dark matter simulations, while
$(\alpha,\beta,\gamma)=(1.5,3,0)$ corresponds to the Burkert cored profile \citep{Bur1995}. Therefore, this dark matter
halo model enables us to explore a wide range of physically plausible dark matter profiles.

In this work, we adopt six parameters $(r_s, \rho_0, \beta_{\rm ani}, \alpha, \beta,\gamma)$ to be determined by fitting to the observed line-of-sight velocity distribution for the GCs in NGC1052-DF2.
In order to set constraints on these parameters and to determine their uncertainties, we utilize Markov Chain Monte Carlo~(MCMC) techniques, based on Bayesian parameter inference, with the standard Metropolis-Hasting algorithm~\citep{Metetal1953,Has1970}\footnote{We take several post-processing steps (burn-in step, the sampling step and length of the chain) to generate independent samples that are insensitive to the initial conditions, and then we obtain the posterior probability distribution function (PDF) of the set of free parameters. By calculating the percentiles of these PDFs, we are able to compute credible intervals for each parameter straightforwardly.}.
For likelihood function, we assume that the line-of-sight velocity distribution is Gaussian and centered on the systemic velocity
of the galaxy $\langle u \rangle$. Given that the total number of tracers is $N$, and the
$i$th tracer has the measured line-of-sight velocity and its observational error $u_i\pm\delta_{u,i}$ at the sky plane coordinates~$(x_i,y_i)$, the
likelihood function is constructed as
\begin{equation}
{\cal L} = \prod^{N}_{i=1}\frac{1}{(2\pi)^{1/2}[(\delta_{u,i})^2 + (\sigma_i)^2]^{1/2}}\exp\Bigl[-\frac{1}{2}\frac{(u_i-\langle u \rangle)^2}{(\delta_{u,i})^2 + (\sigma_i)^2} \Bigr],
\end{equation}
where $\sigma_i$ is the theoretical line-of-sight velocity dispersion at~$(x_i,y_i)$ specified by model
parameters~(as described the above) and derived from the Jeans equations.

Using posterior distribution of each dark halo parameter, we calculate a marginalized dynamical mass at a given radius.
To compare with vD18's results, we estimate the mass within 7.6~kpc, which is the radius of the outermost GC tracer.

%%%%%%%%%%%%%%%%% tables %%%%%%%%%%%%%%%%%%
%%% Table 1 %%%
\begin{table}
	\centering
	\caption{Dynamical mass within 7.6~kpc in the case of the different mass models and adopted stellar scaling radii. The median, 68 and 90 per~cent credible intervals of estimated mass are shown. The Unit of each dynamical mass is $10^{8}M_{\odot}$.}
	\label{table1}
    \scalebox{1.0}[1.2]{
	\begin{tabular}{cccc} % four columns, alignment for each
		\hline\hline
$M_{\rm dm}(<7.6{\rm kpc})$		       &                      &  \multicolumn{2}{c}{Spherical} \\
		       & $R_{\rm half}$ [kpc] &  Isotropic &   Anisotropic   \\
		       \hline
   power-law   & $2.2$ (Stars)        & $0.31^{+0.54+2.83}_{-0.20-0.29}$  & $0.42^{+0.95+3.82}_{-0.33-0.42}$ \\
               & $3.1$ (GCs)          & $0.30^{+0.52+2.51}_{-0.20-0.28}$  & $0.45^{+0.99+3.93}_{-0.35-4.41}$ \\
   S\'ersic    & $2.2$ (Stars)        & $2.97^{+4.33+14.17}_{-1.94-2.77}$ & $4.15^{+6.90+24.31}_{-2.86-4.08}$ \\
               & $3.1$ (GCs)          & $2.60^{+3.76+13.10}_{-1.64-2.37}$ & $3.58^{+6.45+22.47}_{-2.52-3.37}$\\
		\hline\hline
	\end{tabular}
	}
\end{table}

%%% Table 1 %%%
%\begin{table*}
%	\centering
%	\caption{Dynamical mass within 7.6~kpc in the case of the different mass models and adopted stellar scaling radius. The median, 68 and 90 per~cent credible intervals of estimated mass are shown. The Unit of each dynamical mass is $10^{8}M_{\odot}$.}
%	\label{table1}
%    \scalebox{1.0}[1.2]{
%	\begin{tabular}{ccccc} % four columns, alignment for each
%		\hline\hline
%$M_{\rm dm}(<7.6{\rm kpc})$		       &                      &  \multicolumn{2}{c}{Spherical}                  &   Non-spherical \\
%		       & $R_{\rm half}$ [kpc] &  Isotropic &   Anisotropic   &   Anisotropic   \\
%		       \hline
%   power-law   & $2.2$ (Stars)        & $0.31^{+0.54+2.83}_{-0.20-0.29}$  & $0.42^{+0.95+3.82}_{-0.33-0.42}$  & $0.42^{+0.91+3.71}_{-0.31-0.42}$   \\
%               & $3.1$ (GCs)          & $0.30^{+0.52+2.51}_{-0.20-0.28}$  & $0.45^{+0.99+3.93}_{-0.35-4.41}$  &  $0.43^{+0.94+4.21}_{-0.32-0.43}$  \\
%  S\'ersic     & $2.2$ (Stars)        & $2.97^{+4.33+14.17}_{-1.94-2.77}$ & $4.15^{+6.90+24.31}_{-2.86-4.08}$ & $5.21^{+10.96+57.13}_{-3.63-4.93}$ \\
%               & $3.1$ (GCs)          & $2.60^{+3.76+13.10}_{-1.64-2.37}$ & $3.58^{+6.45+22.47}_{-2.52-3.37}$ & $4.49^{+9.19+41.67}_{-3.19-4.25}$  \\
%		\hline\hline
%	\end{tabular}
%	}
%\end{table*}

%%%%%%%%%%%%%%%%%%%%%%%%%%%%%%%%%%%%%%%%%%%%%%%%%%%%%%%%%%%%%%%%%%%%%%%%%%%%%%%5
%%%%%%%%%%%%%%%%%  S3 %%%%%%%%%%%%%%%%%%
%%%%%%%%%%%%%%%%%%%%%%%%%%%%%%%%%%%%%%%%%%%%%%%%%%%%%%%%%%%%%%%%%%%%%%%%%%%%%%%5
\section{Comparison with estimated dynamical masses}
Using the results of the MCMC fitting analysis for the kinematic data of the GC tracers in NGC1052-DF2, 
we estimate the dynamical mass within 7.6~kpc, with marginalizing all dark halo parameters~$(r_s,\rho_0,\alpha,\beta,\gamma)$.
Table~\ref{table1} lists the dynamical masses estimated with the different tracer 
models~(e.g., power-law v.s. S\'ersic profiles, and isotropic v.s. anisotropic velocity 
ellipsoids) and our adopted scale radii of the stellar distributions.
It is found from this result that all masses estimated with the S\'ersic profile are 
systematically more massive than those with the power-law profile.
Also, understandably, considering velocity anisotropy of tracers makes their uncertainties larger.
The reason why using S\'ersic profile makes dynamical mass more massive than power-law may 
be explained as follows:
It is suggested that the shape of tracer density profile have large impact on determination 
of halo density profile~\citep[e.g.,][]{Evaetal2009,Stretal2010,HC2012}.
If a tracer distribution has a flat profile in the central region, the Jeans equation can 
predict a relatively low $\sigma_{l.o.s}$ in the inner region, which can be significantly lower 
than an observed value. In order to match the prediction with the observation in the inner region, 
therefore, the halo modeling method prefers dark matter profiles that have relatively large masses 
in the central region, i.e. it can lead to cuspy density profiles with high $\gamma$, large scale 
radii and/or high scale densities. 
On the other hand, for a steeper tracer density profile in an inner region, it comes out 
in the opposite sense. Namely, a relatively steeper tracer distribution can prefer a less 
massive halo than a flatter tracer distributions. Thus, the halo mass estimation is quite 
sensitive to the assumption of inner tracer profiles, as shown in our result~\footnote{For the case of isotropic velocity ellipsoid, the best-fit values of dark halo parameters, especially~$(r_s,\rho_0,\gamma)$, are $\bigl(\log(r_s)=4.14^{+0.56}_{-0.72},\log(\rho_0)=-3.52^{+1.29}_{-0.79},\gamma=0.57^{+0.40}_{-0.38}\bigr)$ with the S\'ersic and $\bigl(\log(r_s)=3.62^{+0.52}_{-1.12},\log(\rho_0)=-3.60^{+3.01}_{-1.01},\gamma=0.55^{+0.41}_{-0.37}\bigr)$ with the power-law profile, respectively. The values of $r_s$ and $\rho_0$ are in the units of pc and $M_{\odot}$~pc$^{-3}$.}.

\begin{figure}
	% To include a figure from a file named example.*
	% Allowable file formats are eps or ps if compiling using latex
	% or pdf, png, jpg if compiling using pdflatex
	\centering
	\includegraphics[scale=0.50]{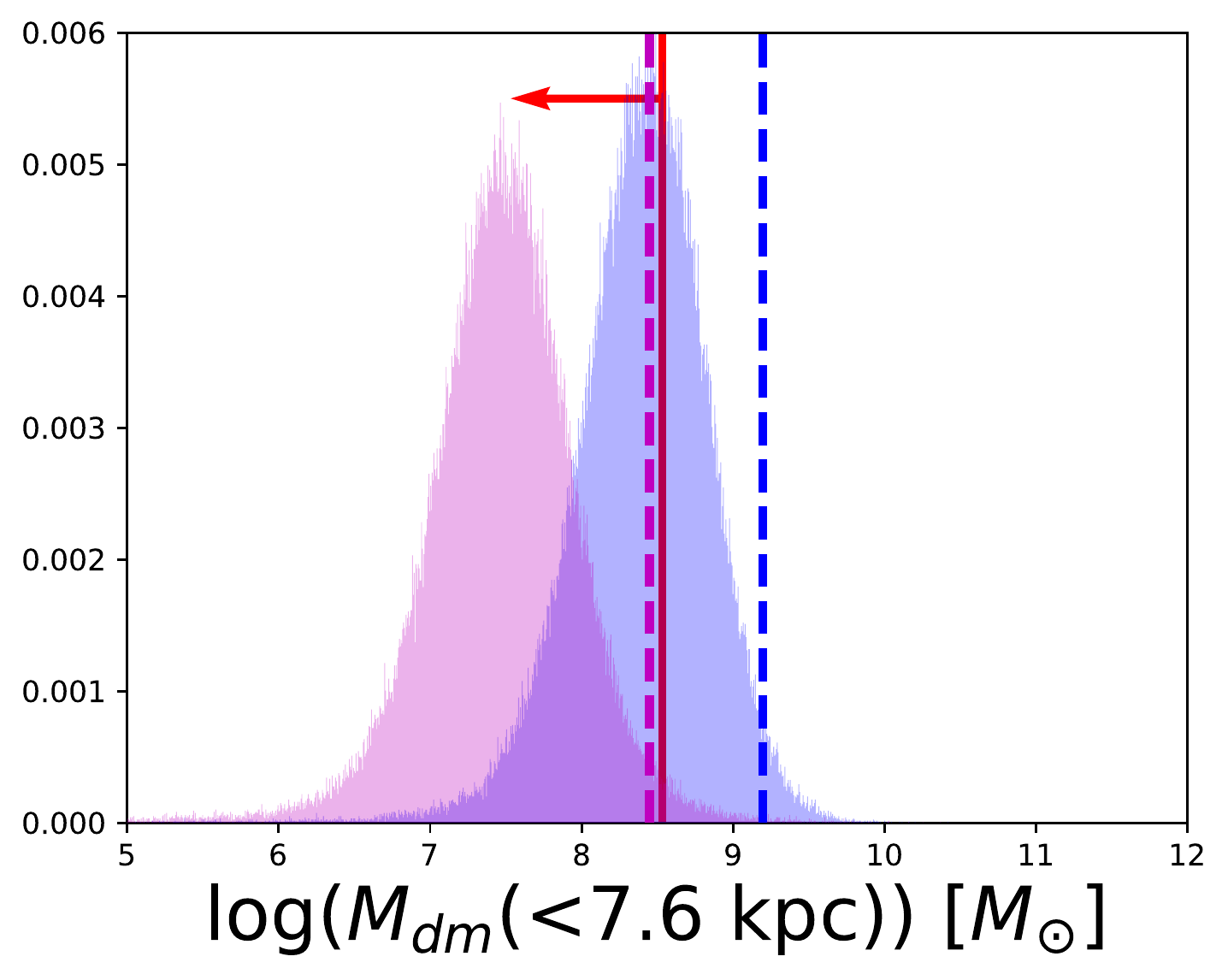}
	\includegraphics[scale=0.50]{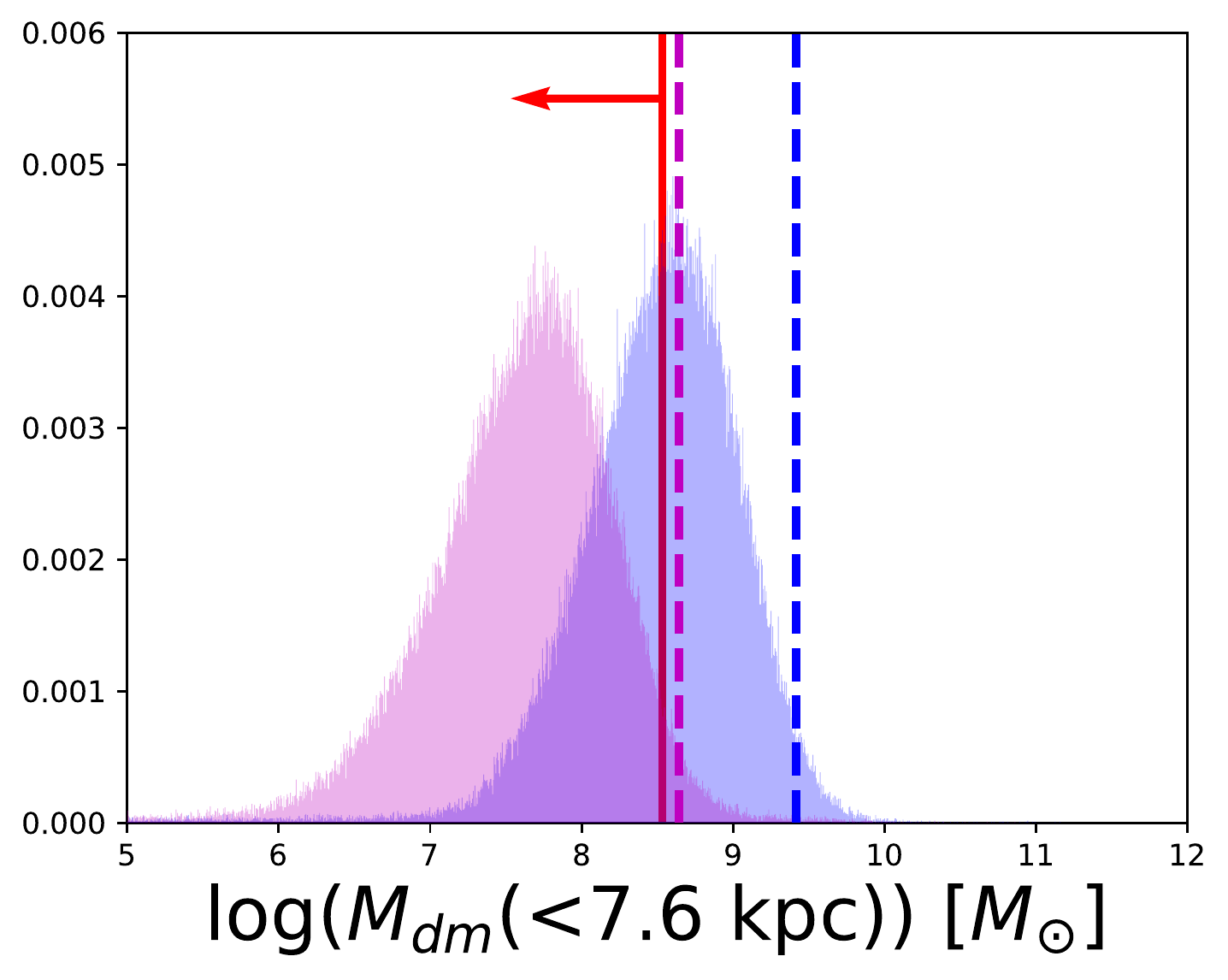}
    \caption{The posterior distribution functions~(PDFs) of marginalized dynamical masses within 7.6~kpc. 
    The magenta and blue PDFs in both panels correspond to the results assuming the power-law and the S\'ersic profiles of the tracer distribution, respectively. 
    Each dashed line indicates 90~per~cent confidential upper limits of the estimated masses, while the solid line with left arrow is the same upper limit of the estimation by vD18.
    Upper panel shows the resultant PDFs with isotropic velocity ellipsoids of tracers, while lower one is those with anisotropic velocity distributions.}
    \label{fig:fig3}
\end{figure}

Figure~\ref{fig:fig3} shows posterior distribution functions~(PDFs) of the marginalized dynamical mass within 7.6~kpc when assuming isotropic~(top panel) and anisotropic velocity distributions~(bottom panel).
Firstly, to compare with the results from vD18, we estimate the mass under the assumption in which the GC tracers have an isotropic velocity ellipsoid, i.e., $\beta_{ani}=0$.
From the top panel, we find that in the case of power-law profile of tracer distribution~(magenta PDF and dashed line), the 90~per~cent upper limit of the estimated mass is almost the same as the result from vD18~(red solid line) which estimates dynamical mass with the same power-law profile.
Thus, we can confirm that our results using Jeans analysis are consistent with those with the MTE if the power-law profile is assumed for the tracers, and thus our method can reproduce the result of vD18.
On the other hand, comparing between the upper limits estimated by vD18 and our S\'ersic tracer density model (blue dashed line), the latter is about one order of magnitude grater than the former.
In addition, the bottom panel shows comparison between mass estimations in the cases where the isotropic assumption is relaxed for the tracer velocities, i.e., $\beta_{ani}\neq0$.
The difference of the upper limits between vD18 and our S\'ersic model still remains in the anisotropic cases. 

We also compare our estimated dynamical mass to those estimated by the other papers~\citep[][]{Lapetal2018,Maretal2018}, which pointed out that there are large uncertainties on dynamical mass estimates of NGC1052-DF2 due to lack of tracer sample and contamination by field GCs.
To this end, applying the mass estimator derived by~\citet{Waletal2009} to the intrinsic velocity dispersions estimated by these studies, we calculate the dynamical masses within 7.6~kpc.
For \citet{Lapetal2018} and \citet{Maretal2018}, the intrinsic velocity dispersions are inferred $\sigma_{\rm int}=11.4^{+5.8}_{-4.5}$~km~s$^{-1}$ and $10.0^{+10.5}_{-3.0}$~km~s$^{-1}$, and then the dynamical masses are $M(<7.6{\rm kpc})=9.85^{+12.55}_{-6.24}\times10^{8}M_{\odot}$ and $7.58^{+24.22}_{-3.97}\times10^{8}M_{\odot}$, respectively.
Comparing them with our estimates shown in Table~\ref{table1}, the masses with their velocity dispersions are as large as those from our analysis with S\'ersic profile, even though their mass estimates are similar to vD18's method. This indicates that small kinematic tracers and the estimates of the intrinsic velocity dispersions of NGC1052-DF2 have a large impact on dynamical mass measurements, thereby implying that in combination with these effects and our dynamical analysis, accuracy of mass estimate of the galaxy could be even worse.
Furthermore, we compare the results of our Jeans models to the mass estimated using the method by~\citet{Erretal2018}.
They adopted the mass estimator, $M_{\rm est}(<1.8R_{\rm half})\approx3.5\times1.8R_{\rm half}\langle\sigma^2_{\rm los}\rangle G^{-1}$, which was introduced by~\citet{AE2012} and~\citet{Cametal2017}, and showed that this estimator would have the minimum uncertainty compared to other estimators.
Using this estimator, we calculate the dynamical masses within the radius of half number of GC tracers~(i.e. $R_{\rm half}=3.1$~kpc) using the intrinsic velocity dispersions, $\sigma_{\rm int}=3.2^{+5.5}_{-3.2}$~km~s$^{-1}$ estimated by vD18 and $11.4^{+5.8}_{-4.5}$~km~s$^{-1}$~estimated by~\citet{Maretal2018}.
As a result, we obtain $M_{\rm est}=0.46^{+3.98}_{-0.46}\times10^{8}M_{\odot}$ for vD18 and $5.90^{+7.50}_{-3.74}\times10^{8}M_{\odot}$ for~\citet{Maretal2018}, and find that there is a little difference between our Jeans analysis and any mass estimators.
This result can confirm that even if we employ the different mass estimators, the estimated mass depends largely on the inferred velocity dispersions.

Finally, we estimate dynamical mass-to-light ratios~($M/L$) assuming the stellar mass of NGC1052-DF2, $M_{\ast}=2.2\times10^8 M_\odot$, derived by vD18 to be fixed.
Table~\ref{table2} shows 90 per~cent confidential upper limits of $M/L$ when we use the dynamical masses within 7.6~kpc.
It is clear that assuming the S\'ersic profile for tracers makes $M/L$ significantly larger than those in the cases assuming the power-law profile.
Moreover, our estimated $M/L\sim7.1$ is similar to those of Fornax and Sculptor classical dwarf spheroidal galaxies~($M/L\sim4.5$ for Fornax and $M/L\sim9.7$ for Sculptor, taken from \citealt{McC2012}). 
Therefore, if the tracer distribution is assumed to follow the S\'ersic profile, NGC1052-DF2 can be considered to be a DM dominant system like the Galactic dwarf galaxies.

%% figure 2

%%% Table 2 %%%
\begin{table}
	\centering
	\caption{The values of 90 per cent confidential upper limits of $M/L$ ratios within 7.6 kpc. The stellar mass of NGC1052-DF2 is fixed to $2.2\times10^8 M_\odot$.}
	\label{table2}
    \scalebox{1.0}[1.0]{
	\begin{tabular}{lc} % four columns, alignment for each
		\hline\hline
	  Models     &      $M/L$     \\
	  \hline
	  Spherical, power-law, isotropy        & $\leq1.28$\\
	  Spherical, S\'ersic, isotropy         & $\leq7.14$\\
	  vD18~(Spherical, power-law, isotropy) & $\leq1.55$\\
%	  \hline
%	  \citet{Lapetal2018}                   & $\leq5.00$\\
%	  \citet{Maretal2018}                   & $\leq8.10$\\
		\hline\hline
	\end{tabular}
	}
\end{table}

%~\textcolor{red}{(but for~\citet{Lapetal2018}, this upper limit of $M/L$ within 3.1~kpc)}

\section{Summary and Conclusions}
NGC1052-DF2 was recently discovered as the dark-matter deficient UDG claimed by vD18.
However, large uncertainties of its dynamical mass estimate have been pointed out in terms of the paucity of 
kinematic sample, statistical methods and distance measurements.

In this work, we argue the effects of the difference in mass models on dynamical mass estimate.
In particular, we focus on the modeling of a tracer density profile and challenge the single power-law profile assumed in vD18. Their MTE modelling based on~\citet{Watetal2010} posits on a power-law profile for a tracer distribution due to requirement from their analytic treatment of the Jeans equation; however it is not justified.
Therefore, we scrutinize the dynamical mass of NGC1052-DF2 with two different tracer 
density models, which are single power-law and S\'ersic profiles.

To begin with, using a simple $\chi^2$ method, we fit these two density profile models to the projected GC tracers cumulative number distribution
to obtain their structural parameters.
As a result, both models are fitted appropriately with little difference in the goodness of fit, due to the sparsity of the data sample.

Using these tracer density models, we derive dynamical masses of the galaxy by comparing the line-of-sight velocity dispersions between the observations of the 10 GC tracers and the analytic predictions computed from the spherical Jeans equation.
In comparison between the estimated masses with the power-law and S\'ersic profiles, we find that the estimated dynamical masses with the S\'ersic model are systematically more massive than those with the power-law model.

A possible reason for this result might be explained as that if a tracer density has a cored profile such as S\'ersic model, the predicted line-of-sight velocity dispersions can decrease within the cored region and be lower than a observed value.
In order to match the predicted value to the observation, the corresponding dark halo parameters such as a scale density and radius tend to become large values that can lead the prediction to a relatively large dynamical mass.
By contrast, in the case for the cusped tracer density, the dark halo parameters tend to become small values that prefer a relatively low dynamical mass.
Consequently, the dynamical mass derived with the cored tracer density tends to be larger than that in the case of the cusped distribution.

We compare our estimated dynamical masses with those calculated by the other studies that pointed out large uncertainties on dynamical mass estimates of NGC1052-DF2 and inferred higher intrinsic velocity dispersions than those from vD18.
Comparing them to our results, their estimated dynamical masses accord roughly with those from our analysis with S\'ersic profiles, and thereby confirming that small tracer sample and the intrinsic velocity dispersion estimation of NGC1052-DF2 have a large impact on dynamical mass measurements.
Also, we confirm that a dynamical mass estimate depends largely on the inferred intrinsic velocity dispersion, irrespective of mass estimator modellings.

Finally, by comparing between the 90~per~cent confidence upper limits of the dynamical masses estimated in this work and vD18, we find two main results.
When assuming the power-law cusped density profile of the tracers, the upper limit of the mass estimated in this case is nearly the same as the result from vD18.
Thus, our results using the Jeans analysis are consistent with those obtained from the MTE modelling used in vD18.
On the other hand, when we adopt the S\'ersic cored density profile for the tracers, the value of upper limit is about one order of magnitude grater than that from vD18. 
Correspondingly, the upper limit of dynamical mass-to-light ratio determined with the S\'ersic density profile is significantly higher than that obtained with the power-law distribution.
Also, this mass-to-light ratio is compatible with those of luminous dwarf galaxies in the Local Group, thereby suggesting that NGC1052-DF2 can still be a dark-matter dominated system, and this galaxy may not be deficient in dark matter.

As our conclusion, dynamical mass estimate of a galaxy with paucity of data is dependent largely on the choice of dynamical 
models, especially tracer distributions, and thus the estimated mass of NGC1052-DF2 is still considered to be highly uncertain.
Therefore, in an attempt to determine robustly dynamical masses of NGC1052-DF2, it is required for dynamical analysis to use kinematic data of stars in the galaxy, and it needs spectroscopic observations; however it is challenging because of the faintness of the galaxy at the moment. 

\section*{Acknowledgements}

We are grateful to the referee for her/his careful reading of our paper and thoughtful comments.
This work was supported in part by the MEXT Grant-in-Aid for Scientific Research on Innovative Areas, No.~18H04359 and No.~18J00277~(for K.H.), and for Young Scientists (B), No.~17K17677 (for S.I.). S.I. was supported by World Premier International Research Center Initiative (WPI), MEXT, Japan and by SPPEXA through JST CREST JPMHCR1414.

%%%%%%%%%%%%%%%%%%%%%%%%%%%%%%%%%%%%%%%%%%%%%%%%%%

%%%%%%%%%%%%%%%%%%%% REFERENCES %%%%%%%%%%%%%%%%%%

%%%%%%%%%%%%%%%%%%%%%%%%%%%%%%%%%%%%%%%%%%%%%%%%%%

% Don't change these lines
\bsp	% typesetting comment
\label{lastpage}
\end{document}